\documentclass[aps,prl,nofootinbib,twocolumn,showpacs,floatfix]{revtex4}%
\usepackage{graphicx}
\usepackage{amssymb}
\usepackage{epstopdf}
\usepackage{graphicx}
\usepackage{amsmath}
\usepackage{amsfonts}
\usepackage{color}%
\providecommand{\U}[1]{\protect\rule{.1in}{.1in}}
\providecommand{\U}[1]{\protect\rule{.1in}{.1in}}
\begin{document}
\title{Melting the Superconducting State in the Electron Doped Cuprate Pr$_{1.85}%
$Ce$_{0.15}$CuO$_{4-\delta}$ with Intense near-infrared and Terahertz Pulses}
\author{M. Beck,$^{1}$ M. Klammer,$^{1}$ I. Rousseau,$^{1}$ M. Obergfell,$^{1,2}$ P.
Leiderer,$^{1}$ M. Helm,$^{3,4}$ V.V. Kabanov,$^{5}$ I. Diamant,$^{6}$ A.
Rabinowicz,$^{6}$ Y. Dagan,$^{6}$ and J. Demsar$^{1,2}$}
\affiliation{$^{1}$Dept. of Physics and Center for Applied Photonics, Univ. of Konstanz,
D-78457, Germany}
\affiliation{$^{2}$Institute of Physics, Johannes Gutenberg-University Mainz, 55128 Mainz,
Germany }
\affiliation{$^{3}$Institute of Ion Beam Physics and Materials Research, Helmholtz-Zentrum
Dresden-Rossendorf, P.O. Box 510119, 01314 Dresden, Germany}
\affiliation{$^{4}$Technische Universit\"{a}t Dresden, 01062 Dresden, Germany}
\affiliation{$^{5}$Jozef Stefan Institute, Ljubljana, SI-1000, Slovenia}
\affiliation{$^{6}$Raymond and Beverly Sackler School of Physics and Astronomy, Tel Aviv
University, Tel Aviv, 69978, Israel}


\begin{abstract}
We studied the superconducting (SC) state depletion process in an electron
doped cuprate Pr$_{1.85}$Ce$_{0.15}$CuO$_{4-\delta}$ by pumping with
near-infrared (NIR) and narrow-band THz pulses. When pumping with THz pulses
tuned just above the SC gap, we find the absorbed energy density required to
deplete superconductivity, $A_{dep}$, matches the thermodynamic condensation
energy. Contrary, by NIR pumping $A_{dep}$ is an order of magnitude higher,
despite the fact that the SC gap is much smaller than the energy of relevant
bosonic excitations. The result implies that only a small subset of bosons
contribute to pairing.

\end{abstract}
\maketitle

The quest for a pairing boson in cuprate high-temperature superconductors has
been one of the key topics of solid state physics ever since the discovery of
superconductivity in the cuprates. Recently, numerous femtosecond (fs)
real-time studies of carrier dynamics in high-T$_{c}$ superconductors have
been performed aiming to find the coupling strengths between the electrons and
other degrees of freedom (high and low frequency phonons, spin fluctuations,
electronic
continuum)\cite{Perfetti,Kusar,Gadermaier,DelConte,Mansart,StojchevskaPni,Bovensiepen,Lanzara}%
. In this approach, fs optical pulses are used to excite the electronic
system, while the resulting dynamics are probed by measuring the changes in
optical constants \cite{Kusar,Gadermaier,DelConte,Mansart,StojchevskaPni} or
the electronic distribution near the Fermi energy
\cite{Perfetti,Bovensiepen,Lanzara}. To connect the measured relaxation
timescales to the electron-boson coupling strengths, the multi-temperature
models are commonly used \cite{Kaganov,AllenEx}. These are based on the
premise that the electron-electron (\textit{e-e}) thermalization is much
faster than the electron-boson relaxation. While these models are commonly
used to extract e.g. the electron-phonon (\textit{e-ph}) coupling strengths,
numerous inconsistencies have been noted (even for the case of simple metals)
\cite{Fann,Groeneveld,KabAlex,BookChapter}. An alternative time-domain
approach, based on the dynamics in the superconducting state, has been put
forward \cite{RT,Beck1}. Under the assumption that the absorbed optical energy
is distributed between quasiparticles and high frequency ($\hbar\omega
>2\Delta$) bosons on the sub-picosecond timescale, and taking into account the
nonlinearity of relaxation processes (pairwise recombination of
quasiparticles), the electron-boson coupling strength is determined by
studying the excitation density dependence of the Cooper pair-breaking process
\cite{RT,Beck1}. While this approach has been successfully applied to
conventional superconductors \cite{Beck1,MgB2}, the results on cuprates show
that the energy density required to suppress superconductivity exceeds the
thermodynamic condensation energy, $E_{c}$, by an order of magnitude
\cite{Kusar,Giannetti,Beyer,Stojcevska}. Therefore, the assumption that the
absorbed energy is distributed between QPs and the coupled high frequency
bosons fails. Considering the possible energy relaxation pathways, this
discrepancy in the hole-doped high-T$_{c}$ cuprates has been attributed to the
fact that the superconducting gap, 2$\Delta$, lies well in the range of
optical phonons \cite{Beyer}. It has been argued that $\approx90$ $\%$ of the
absorbed energy is directly released to $\hbar\omega<2\Delta$ modes via rapid
\textit{e-ph} scattering (mainly to sub-gap zone center optical phonons and
zone-edge acoustic phonons) and only $\approx10$ $\%$ is available for
condensate quenching \cite{Beyer}. Indeed, it has been shown that in YBCO the
rapid \textit{e-ph} transfer gives rise to a rapid heating of specific phonons
on the timescale of $\approx100$ fs \cite{Pashkin}. Studying cuprate
superconductors with $2\Delta$ far below the energy of optical phonons can put
this argument to the test.

In this Letter we present a systematic study of light induced quenching of
superconductivity in an e-doped cuprate \cite{RMP} superconductor
$\text{Pr}_{1.85}\text{Ce}_{0.15}\text{CuO}_{4-\delta}$ (PCCO) at optimal
doping. We used near-infrared ($\lambda=800$ nm) as well as narrow-band THz
($\lambda=144$ $\mu$m) excitation, while probing the superconducting gap
dynamics with THz probe pulses. In PCCO $2\Delta\approx7$ meV
\cite{Dagan1,Dagan2,Zimmers,Homes}, well below the acoustic phonon cut-off
frequency of $\approx20$ meV \cite{PDOS}, as well as the energy of the
collective electronic mode of $\approx11\pm2$ meV \cite{MagMode1,MagMode2}.
Like in NbN \cite{Beck1}, we expected the absorbed energy density required to
deplete superconductivity, $A_{dep}$, to be in the low temperature limit the
same in the two configurations, with $A_{dep}\approx E_{c}$. We demonstrate
that this is not the case, suggesting that the Eliashberg electron-boson
coupling function in cuprates depends strongly on\ the electron energy, unlike
to what is commonly assumed \cite{AllenEx,Allen}.

Optimally doped c-axis oriented PCCO thin films with a thickness $d=60$ nm
were epitaxially grown on LaSrGaO$_{4}$ (001) (LSGO) substrates using pulsed
laser deposition \cite{Maiser}. Inductive measurements of the samples yield a
$T_{c}\approx21$ K. The broadband linear and time-resolved THz spectroscopy
were performed on the set-up built around a 250 kHz amplified Ti:sapphire
laser system and utilizing large area interdigitated photoconductive emitter
for the generation of THz pulses \cite{BeckOptExp}. Narrow band THz pumping
experiments were performed at the free electron laser (FEL) facility at the
Helmholtz-Zentrum Dresden-Rossendorf. Here, intense narrowband (spectral width
$\approx30$ GHz, pulse length $\tau_{\text{FEL}}\approx18$ ps) THz pulses at
$\nu_{\text{FEL}}=2.08$ THz, slightly above the low-temperature gap frequency
2$\Delta/h\approx1.7$ THz \cite{Dagan1,Dagan2}, were used as both pump and
probe sources using the configuration described in Ref.\cite{Beck2}.

The equilibrium THz conductivity studies were performed by recording the THz
electric fields transmitted through the sample (film on substrate),
$E_{tr}(t^{\prime})$, and the reference (bare substrate), $E_{re}(t^{\prime}%
)$, using the Pockels effect in GaP. The bandwidth was limited to below 2.5
THz by the TO-Phonon of the LSGO. The complex optical conductivity
$\sigma(\omega)=\sigma_{1}(\omega)+i\sigma_{2}(\omega)$ was obtained \cite{SI}
using the Fresnel equations. The normal state $\sigma(\omega)$ can be well
approximated by the Drude model, giving the plasma frequency $\nu_{p}=190$ THz
and the scattering rate $\tau^{-1}=2.2$ THz, in good agreement with infrared
studies on thick films \cite{Zimmers,Homes}. To estimate the magnitude of
2$\Delta$, we follow the approach of Ref.\cite{Homes}, where 2$\Delta$ was
extracted from the reflectivity, by reading out the position in the
reflectance spectra, below which the reflectivity starts to rise steeply above
its normal state value. To do so, we used the measured $\sigma(\omega,T)$,
calculated the corresponding bulk reflectivity $R(\omega,T)$, and plot
$R(\omega,T)-$ $R(\omega,$30 K$)$, as shown in Fig. 1(d). It follows that the
maximum gap frequency $2\Delta/\hbar\approx1.7$ THz ($\Delta=3.5$ meV), in
good agreement with other studies on c-axis films \cite{Dagan1,Homes}. Unlike
in the BCS case \cite{Beck1}, $2\Delta$ displays only weak T-dependence.

\begin{figure}[ptb]
\centerline{\includegraphics[width=86mm]{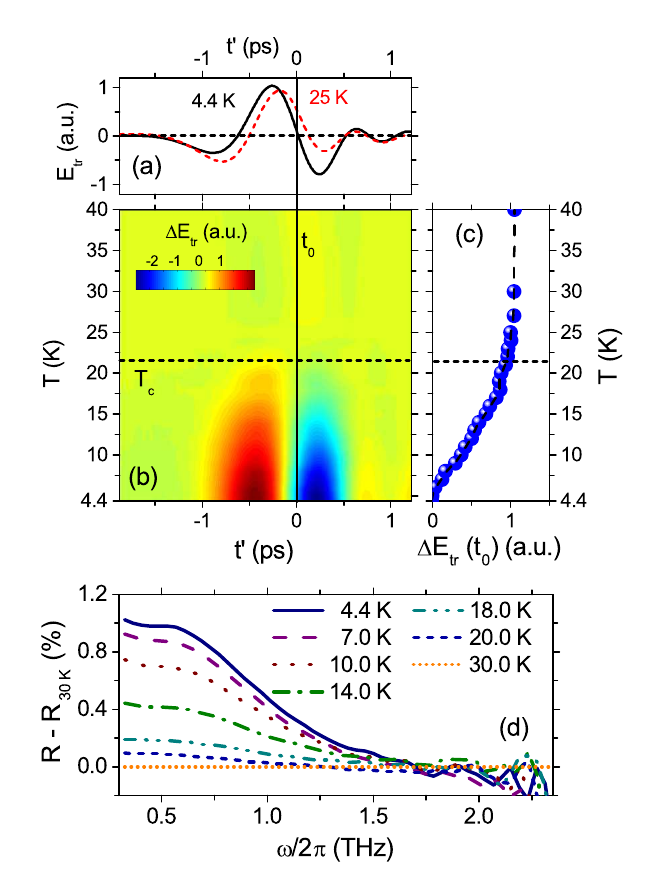}}\caption{(color online)
Temperature dependence of transmitted THz electric field transients. Panel (a)
shows E$_{tr}(t\prime)$ in the normal and superconducting states. The
pronounced phase shift is characteristic of the inductive response below
T$_{c}$. Panel (b) presents $\Delta$E$_{tr}$(T,t$^{\prime}$) = E$_{tr}%
$(T,t$^{\prime}$) - E$_{tr}$(25 K,t$^{\prime}$), while (c) shows the
temperature dependence of $\Delta$E$_{tr}$(T,t$_{0}$) = E$_{tr}$(T,t$_{0}$) -
E$_{tr}$(4.4 K,t$_{0}$). Panel (d) presents the temperature dependence of the
bulk reflectivity extracted from $\sigma\left(  \omega\right)  $ \cite{SI}.}%
\end{figure}

In optical-pump -- THz probe experiments, the film is excited by a 50 fs
near-infrared (NIR) pump pulse centered at 800 nm. The transient spectral
conductivity $\sigma(\omega,t_{d})$ was measured as a function of time delay
$t_{d}$ between the pump and the THz probe pulse. The quenching of the SC
state is found to proceed on a timescale of several ps, while the SC state
recovery was on a timescale of 100's of ps. For time delays $t_{d}$ longer
than the characteristic e-e and e-ph thermalization times (both on the order
of 1 ps) we find that the measured $\sigma(\omega,t_{d})$ can be matched to
the equilibrium $\sigma(\omega)$ recorded at a specific temperature $T^{\ast}$
\cite{SI}, i.e. $\sigma(\omega,t_{d})\approx\sigma(\omega,T^{\ast})$. This is
consistent with the so called T$^{\ast}$-model, where in non-equilibrium the
population of QPs, Cooper pairs and high-frequency ($\hbar\omega>2\Delta$)
bosons are in a quasi-equilibrium at a temperature T$^{\ast}$, which is larger
than the base temperature T of $\hbar\omega<2\Delta$ modes.

To study the excitation density dependence of the SC state dynamics over large
range of excitation densities we recorded the induced changes in the
transmitted THz electric \ field at a fixed $t^{\prime}=t_{0}$ as a function
of $t_{d}$, $\Delta E_{tr}(t_{0},t_{d})$, as in Ref. \cite{Beck1}. As shown in
Fig. 1(a)-(c) the transmitted electric field $E_{tr}(t^{\prime})$ depends
strongly on temperature. In particular,\ for the chosen $t_{0}$ - see Fig. 1 -
$E_{tr}(t_{0},T)$ shows a linear T-dependence over a large T-range below
T$_{c}$.

\begin{figure}[ptb]
\centerline{\includegraphics[width=86mm]{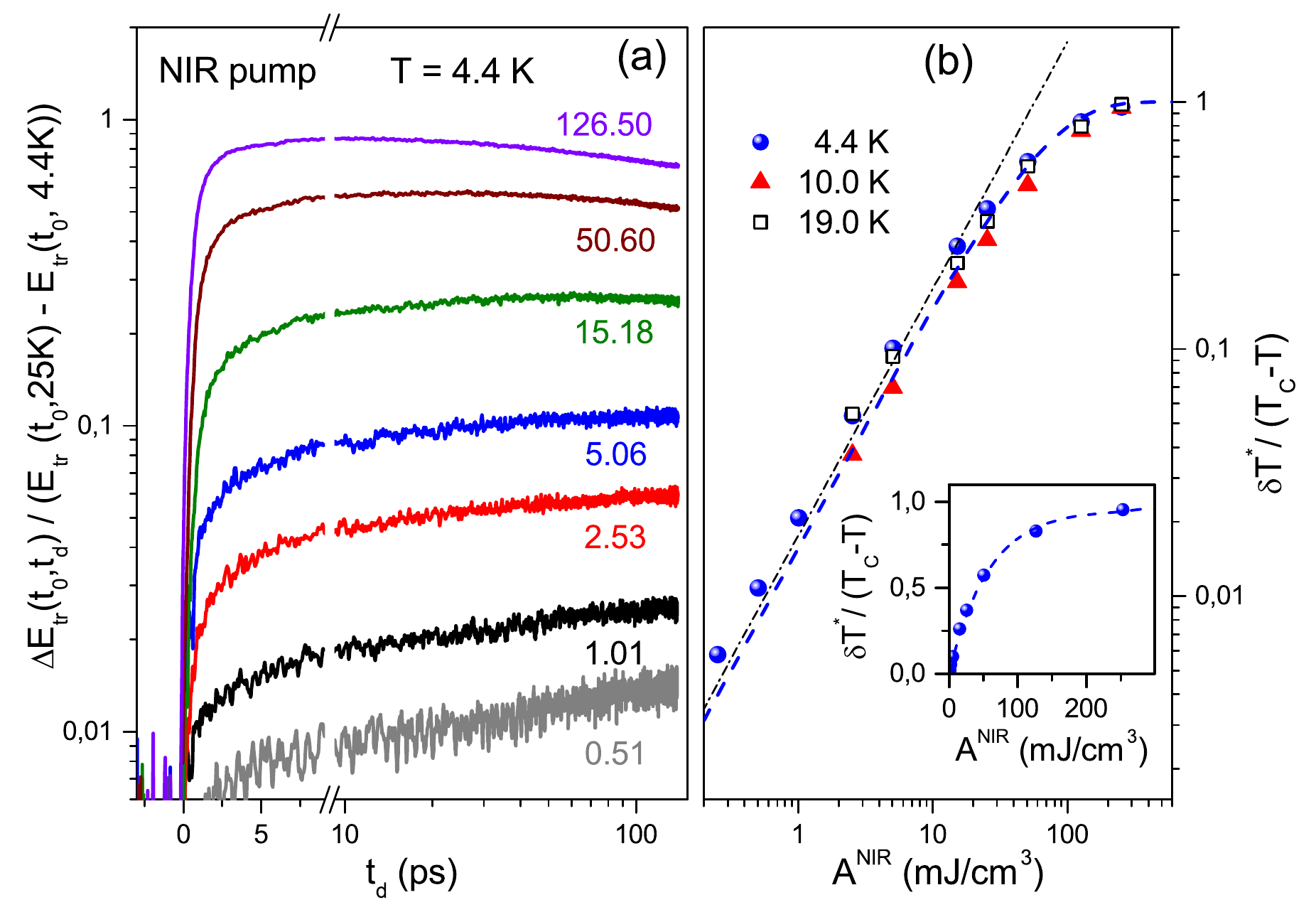}}\caption{(color online) (a)
Transient changes of the transmitted THz electric field $\Delta E_{tr}%
(t_{0},t_{d})$, recorded at $4.4$ K for different absorbed energy densities
$A^{\text{{\protect\tiny NIR}}}$ in mJ/cm$^{3}$. The curves are normalized to
the change in $E_{tr}\left(  t_{0}\right)  $ between 4.4 K and 25 K (above
T$_{c}$). (b) Maximum increase in effective temperature, $\delta$T$^{\ast}$,
as a function of $A^{\text{{\protect\tiny NIR}}}$ for three base temperatures,
normalized to the respective maximum change in T$^{\ast}$, i.e. T$_{c}-$T. The
dashed (blue) line presents the saturation model fit, while the dotted line
presents the linear extrapolation. Inset presents the 4.4 K data on a linear
scale.}%
\end{figure}

Figure 2(a) shows the recorded $\Delta E_{tr}(t_{0},t_{d})$ transients for
$T=4.4$ K and a set of absorbed optical energy densities $A^{\text{{\tiny NIR}%
}}$ (in mJ/cm$^{3}$). Here, $A^{\text{{\tiny NIR}}}$ is extracted from the
incoming laser pulse fluence and the absorption coefficient of PCCO at 800 nm,
which we have measured. It follows from the data that both, SC state depletion
and recovery depend on excitation density. Similar observations have been made
on MgB$_{2}$ \cite{MgB2} and NbN \cite{Beck1}, and could be well accounted by
the phenomenological Rothwarf-Taylor model \cite{RT}. In particular, the slow
timescale for depletion of the SC state (for lowest excitation densities
several 10s of ps!) and its excitation density dependence can be attributed to
Cooper pair-breaking by $\hbar\omega>2\Delta$ bosons generated during the
relaxation of high energy quasiparticles towards the gap \cite{RT,Beck1,MgB2}.

Here, we focus on the energetics of the SC state depletion. We chose $\Delta
E_{tr}(t_{0},t_{d}=30$ ps$)$ to determine the change in effective temperature
T$^{\ast}$ as a function of excitation density. The link to T$^{\ast}$ is
provided by the T-dependence of $\Delta E_{tr}(t_{0},T)$ in equilibrium shown
in Fig. 1(c); we used linear interpolation for low excitation densities. The
relative change in T$^{\ast}$, $\delta$T$^{\ast}/($T$_{c}-$T$)$, as a function
of $A^{\text{{\tiny NIR}}}$ is shown in Fig. 2(b). For low
$A^{\text{{\tiny NIR}}}$ the induced changes scale with $A^{\text{{\tiny NIR}%
}}$, yet showing the expected saturation at high densities. By applying a
simple saturation model fit, where $\frac{\delta T^{\ast}}{T_{c}%
-T}=1-\mathrm{exp}(-A^{\text{{\tiny NIR}}}/A_{dep}^{\text{{\tiny NIR}}})$, we
extract the absorbed energy density required for depleting the SC state,
$A_{dep}^{\text{{\tiny NIR}}}$. At 4.4 K we find $A_{dep}^{\text{{\tiny NIR}}%
}\approx60$ mJ/cm$^{3}$, comparable to the value obtained on Nd$_{1.85}%
\text{Ce}_{0.15}\text{CuO}_{4+\delta}$ \cite{Hinton}. This energy is 6 times
the superconducting state condensation energy of PCCO, $E_{c}\simeq10$
mJ/cm$^{3}$ \cite{Balci}, following the trend of hole-doped cuprates where
$A_{dep}^{\text{{\tiny NIR}}}\gg E_{c}$ \cite{Kusar,Beyer,Stojcevska,Hinton}.
We note that $A_{dep}^{\text{{\tiny NIR}}}$ is substantially lower than the
energy needed to thermally suppress superconductivity given by $E_{th}=%
{\textstyle\int\nolimits_{4.4K}^{21K}}
C_{p}\left(  T\right)  dT\simeq250$ mJcm$^{-3}$, where $C_{p}\left(  T\right)
$ is the total specific heat \cite{Balci}.

We further studied the SC state quenching in PCCO using intense THz pulses.
This way the electronic system is excited directly, with quasiparticles having
small excess energy. A single-color pump probe scheme was used \cite{Beck2}.
The T-dependence of transmission intensity ($Tr$) in equilibrium is shown in
inset to Fig. 3(a). The measurements agree with the transfer-matrix method
(TMM) calculation based on the optical conductivity data obtained by broadband
THz spectroscopy \cite{SI}.

\begin{figure}[ptb]
\centerline{\includegraphics[width=86mm]{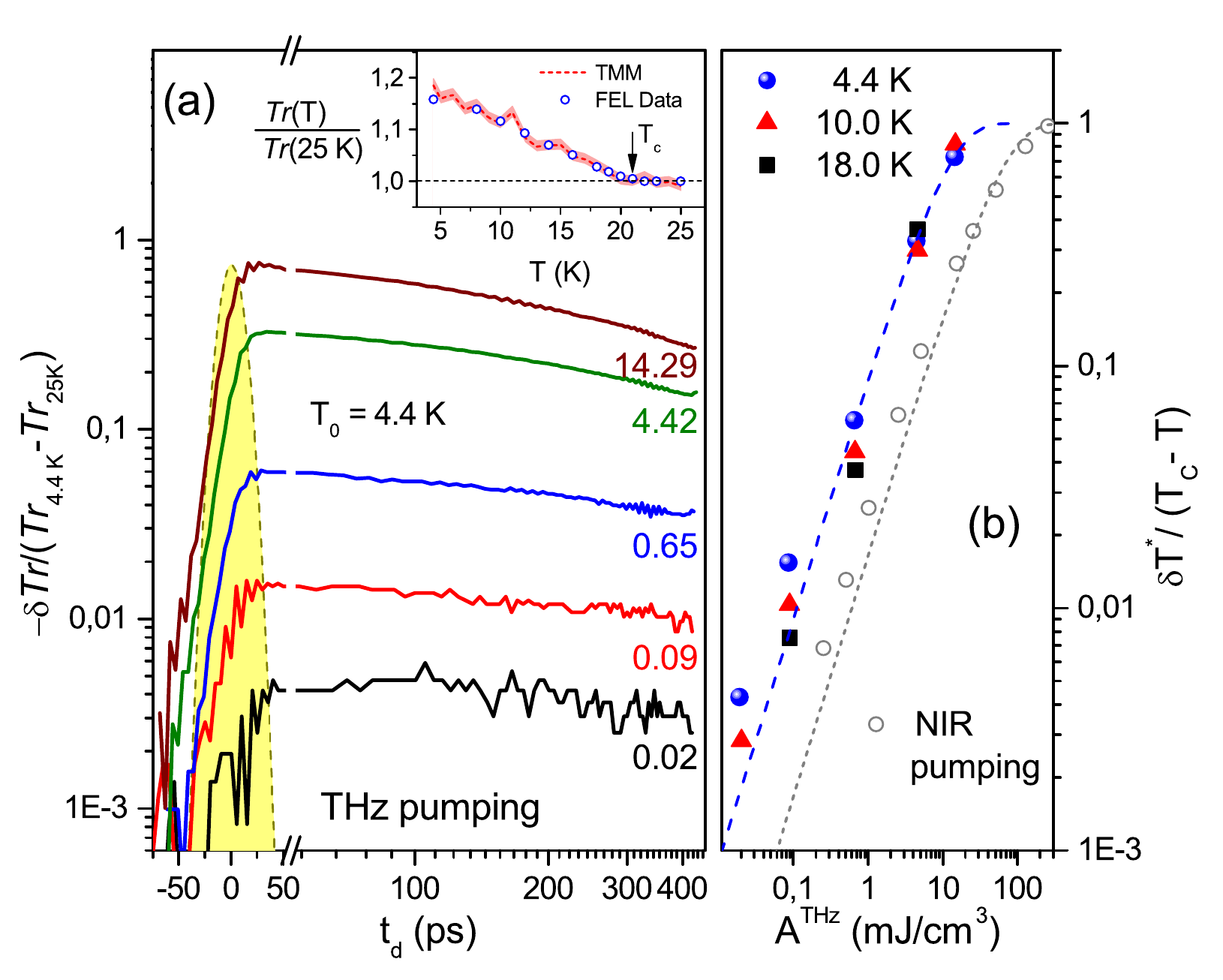}}\caption{(color online) (a)
The changes in the THz transmission intensity $\delta Tr$ after
photoexcitation with intense THz pulses at $T=4.4$ K (absorbed energy
densities $A^{\text{{\protect\tiny THz}}}$ in mJ/cm$^{3}$ are listed). The
shaded area presents the cross-correlation of narrow-band THz pulses. Inset
shows the T-dependence of equilibrium $Tr$. The measured transmission in the
FEL set-up matches well the TMM calculation based on the measured
$\sigma\left(  \omega\right)  $. (b) The change in effective temperature
$\delta T^{\ast}$ as a function of $A^{\text{{\protect\tiny THz}}}$ for three
base temperatures. Dashed blue line is the fit with simple saturation model.
Open (grey) circles and the dotted (grey) line present the data obtained by
NIR pumping for comparison.}%
\end{figure}

Figure 3(a) presents the time-evolution of the normalized pump-induced changes
in transmission ($\delta Tr$) recorded for several $A^{thz}$. The shaded area
corresponds to the intensity cross-correlation of THz pulses. Unlike in the
NIR-pump experiments, no SC state suppression dynamics beyond the THz pump
pulse duration is observed, despite the fact that the lowest excitation
densities were an order of magnitude lower than in the NIR pump study. This
clearly demonstrates different excitation mechanisms. While by pumping with
THz photons with $\nu_{\text{FEL}}>2\Delta/\hbar$ Cooper pairs are broken
directly, by NIR pumping Cooper pairs are broken mainly by $\hbar
\omega>2\Delta$ bosons generated during the cascade of high energy
quasiparticles towards the gap edge.

Assuming the T$^{\ast}$-approximation as for NIR pumping, we convert the
maximum induced changes in $Tr$ into changes of T$^{\ast}$, using the
calibration curve in inset to Fig. 3(a). Figure 3(b) summarizes the results
for three different base temperatures, indicating similar values of
$A_{dep}^{\text{{\tiny THz}}}$. Contrary to $A_{dep}^{\text{{\tiny NIR}}}$,
$A_{dep}^{\text{{\tiny THz}}}\left(  4.4\text{ K}\right)  \approx11$
mJ/cm$^{3}$ is within errorbars identical to $E_{c}$. This implies that almost
all of the deposited energy is directly used for condensate depletion and only
a miniscule amount of energy is transferred into the bosonic system. Since the
recovery dynamics is identical for the two excitation processes, and can be
attributed to the boson-bottleneck scenario, some energy has to be transferred
to the $\hbar\omega>2\Delta$ bosonic modes. However, due to the detailed
balance \cite{RT,MgB2} between the QP density, $n_{qp}\varpropto\exp\left(
-\Delta/k_{B}T\right)  $, and the $\hbar\omega>2\Delta$ boson density,
$n_{b}\varpropto\exp\left(  -2\Delta/k_{B}T\right)  $, where $n_{b}\varpropto
n_{qp}^{2}$ , the energy stored in the latter is negligible for $\Delta
>k_{B}T$.

\begin{figure}[ptb]
\centerline{\includegraphics[width=86mm]{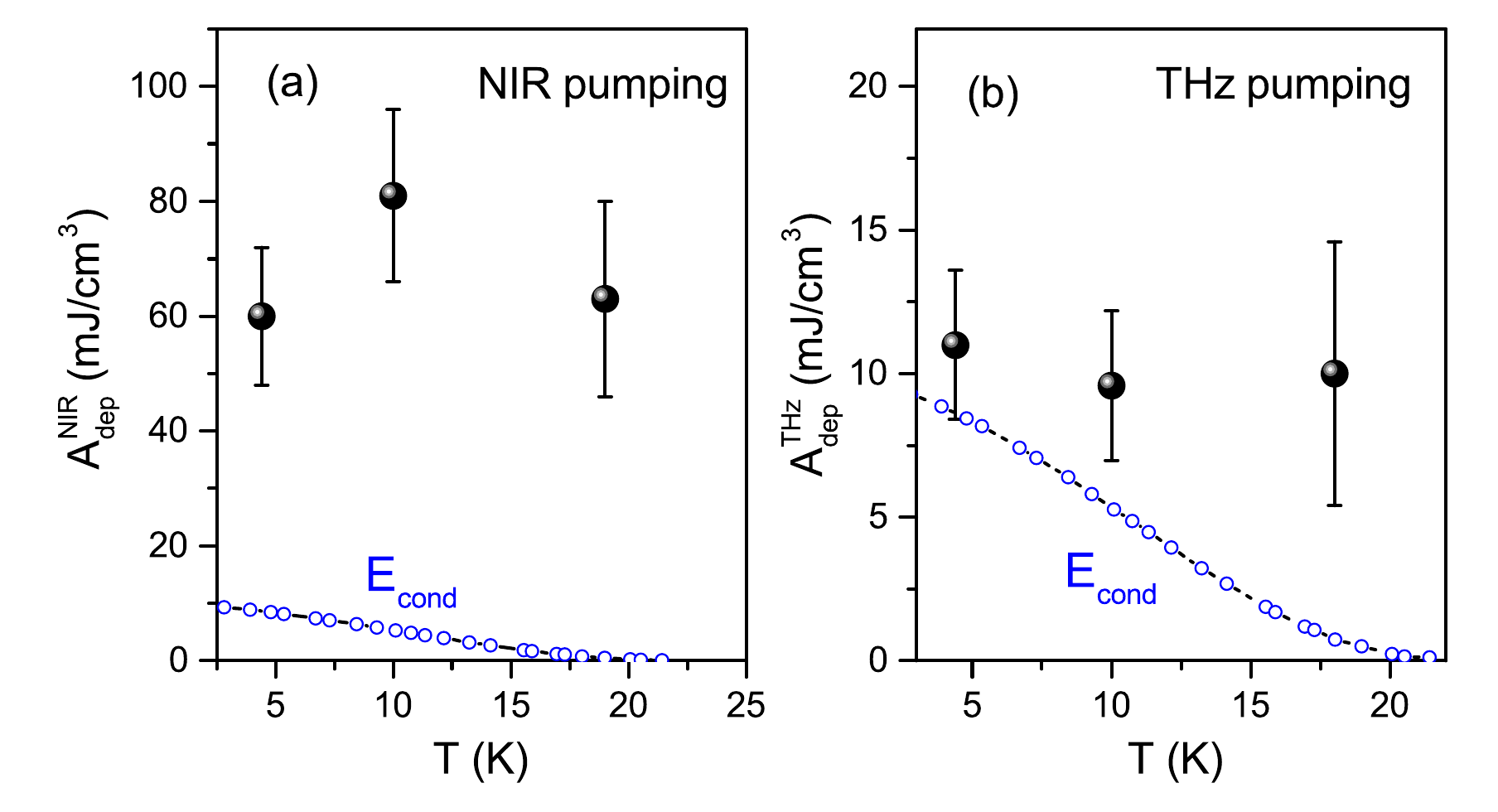}}\caption{Comparison of the
absorbed optical energy density required for superconducting state depletion
for (a) pumping with NIR (1.5 eV) and (b) THz (8.6 meV) pulses. }%
\end{figure}

Figure 4 summarizes the dependence of $A_{dep}$ on the pump photon energy and
T. The observation that the low-T value of $A_{dep}^{\text{{\tiny THz}}}$
matches $E_{c}$ is consistent with expectations. Far more surprising is the
result of the NIR pumping. Considering the fast condensate depletion times and
specific heats of different bosonic subsystems (phonons, spin fluctuations),
the observation in hole doped cuprates that $A_{dep}^{\text{{\tiny NIR}}}\gg
E_{c}$ was explained by the dominant \textit{e-ph} (pairing) interaction
\cite{Kusar}. In this scenario $\hbar\omega>2\Delta$ phonons couple to the
condensate via pair-breaking (and re-pairing), while a large amount of the
absorbed energy is rapidly transferred to optical phonons with $\hbar
\omega<2\Delta$ \cite{Beyer,Stojcevska}. These act as an effective heat sink,
yet lack the energy for breaking up of Cooper pairs. According to this
scenario, $A_{dep}^{\text{{\tiny NIR}}}\approx E_{c}$ for superconductors with
small $\Delta$ \cite{Beyer,Stojcevska}, as shown for NbN \cite{Beck1} and
ferropnictides \cite{Stojcevska}. Since in PCCO $2\Delta$ $\approx7$ meV is
far below the acoustic phonon cut-off energy of $\approx20$ meV \cite{PDOS},
this interpretation is hereby challenged. Since $A_{dep}^{\text{{\tiny NIR}}%
}\ll E_{th}$ we can exclude $\hbar\omega<2\Delta$ acoustic phonons as an
effective energy sink during the initial \textit{e-ph} cascade. Consequently,
our results suggest a rapid \textit{electron-boson} energy transfer following
NIR pumping, yet \textit{only selected modes couple to the condensate}. We
should note that no significant variation of $A_{dep}$ on base temperature is
observed. This is not surprising for NIR pumping, where $A_{dep}%
^{\text{{\tiny NIR}}}\gg E_{c}$, and $\leq20\%$ of absorbed energy is used for
condensate depletion. The fact that similar is observed by THz pumping is
puzzling. We speculate that this may be linked to the enhancement of SC due to
nonequilibrium electron distribution as observed in NbN \cite{Beck2}.

The observation that in PCCO in the low-T limit $A_{dep}^{\text{{\tiny NIR}}%
}\gg A_{dep}^{\text{{\tiny THz}}}\approx E_{c}$ implies, quite generally, that
the Eliashberg coupling function strongly depends on the electron energy. The
electron-boson spectral function $\alpha^{2}F(\varepsilon,\varepsilon^{\prime
},\Omega)$, where $\varepsilon$ and $\varepsilon^{\prime}$ are the electronic
and $\Omega$ the bosonic energy, is commonly approximated by $\alpha
^{2}F(\varepsilon_{F},\varepsilon_{F},\Omega)=\alpha^{2}F(\Omega)$, since it
is customary to assume that its variation on $\varepsilon$ and $\varepsilon
^{\prime}$ on the energy scale smaller than the electronic bandwidth can be
neglected \cite{Allen}. Our result, that by NIR pumping most of the energy is
transferred to bosonic excitations, which at the same time do not couple to
the condensate, suggests the failure of the above assumption in cuprates. This
implies that $\alpha^{2}F$ changes dramatically on the energy scale of (a few)
100 meV.

Considering the possible scenarios of superconductivity being mediated by the
phonons or magnetic excitations, the results suggest that high energy
electrons strongly couple to phonons (or magnetic modes), while the situation
is reversed at low energies. Here, by high energy we refer to $\approx100$
meV, since for energies of the order of the NIR photon energy the e-e
scattering dominates the e-boson scattering thereby substantially reducing the
average electron excess energy on the timescale of a few fs \cite{Petek}.
Presuming the situation is similar in both electron and hole doped cuprates,
and taking into account the result that NIR pumping in YBCO\ results in a
rapid e-ph energy transfer \cite{Pashkin}, this result may suggest that
pairing in cuprates is mediated by magnetic excitations. Alternatively, we
could assume that high energy electrons emit magnetic excitation on the fs
timescale \cite{DalConte}. If so, these nonequilibrium magnetic excitations
are almost uncoupled from the condensate and therefore do not act as pair-breakers.

Note that the proposed scenario is based on a very simple observation that the
excitations created by the high energy electrons are poorly coupled to the
condensate. Therefore the relaxation times due to this coupling is longer than
the anharmonic decay of these excitations. As a result Rothwarf and Taylor
bottleneck \cite{RT} is not operational for these excitations.

In summary, we present photoinduced THz conductivity dynamics in the
electron-doped cuprate PCCO, focusing on melting the SC state with both NIR
and THz pulses. The absorbed energy density required for SC state depletion,
$A_{dep}$, was found to match the condensation energy when pumping with narrow
band THz pulses. With NIR excitation, however, $A_{dep}^{\text{{\tiny NIR}}%
}\gg E_{c}$, despite the fact that $2\Delta$ is small compared to relevant
bosonic energy scales. The data imply that following NIR pumping a rapid
\textit{electron-boson} energy transfer takes place, yet only selected bosonic
modes (e.g. antiferromagnetic fluctuations, or specific lattice modes) do
couple to the condensate. Further systematic studies, where excitation energy
is varied over a large energy range, and supported by theoretical modelling
are clearly required in cuprates, as well as in other systems with competing orders.

\begin{acknowledgments}
This work was supported by the German Israeli DIP project No. 563363,
Alexander von Humboldt Foundation, Kurt Lion Foundation, Zukunftskolleg and
Center for Applied Photonics at the University of Konstanz. We thank P. Michel
and the FELBE team for their dedicated support. Y.D. acknowledes support from
the Israel Science Foundation under grant 569/13.
\end{acknowledgments}

\end{document}